\documentclass[a4paper,fleqn,usenatbib]{mnras}

\usepackage{upgreek}
\usepackage[T1]{fontenc}
\usepackage{graphicx}	% Including figure files
\usepackage{amsmath}	% Advanced maths commands
\usepackage{amssymb}	% Extra maths symbols

\usepackage{cancel}
\newcommand{\smccep}{}
\newcommand{\lmccep}{}
\newcommand{\sn}{\ifmmode {\rm S/N}\else${\rm S/N}$\fi}
\newcommand{\fx}{\ifmmode \nu_{\rm x}\else$\nu_{\rm x}$\fi}
\newcommand{\Pm}{\ifmmode P_{\rm m}\else$P_{\rm m}$\fi}
\newcommand{\Pf}{\ifmmode P_{\rm 0}\else$P_{\rm 0}$\fi}
\newcommand{\fF}{\ifmmode \nu_{\rm 0}\else$\nu_{\rm 0}$\fi}
\renewcommand{\fm}{\ifmmode \nu_{\rm m}\else$\nu_{\rm m}$\fi}
\newcommand{\fO}{\ifmmode \nu_{\rm 1}\else$\nu_{\rm 1}$\fi}
\title[Modulation in classical Cepheids]{Unstable standard candles. Periodic light curve modulation in fundamental mode classical Cepheids.}

\author[Smolec]
{
R. Smolec,$^{1}$\thanks{E-mail: smolec@camk.edu.pl}
\\
$^{1}$Nicolaus Copernicus Astronomical Center
              of the Polish Academy of Sciences,
              Bartycka 18, PL-00-716 Warszawa, Poland\\
}

\date{Accepted XXX. Received YYY; in original form ZZZ}

\pubyear{2017}

\begin{document}
\label{firstpage}
\pagerange{\pageref{firstpage}--\pageref{lastpage}}
\maketitle

\begin{abstract}
We report the discovery of periodic modulation of pulsation in 51 fundamental mode classical Cepheids of the Magellanic Clouds observed by the Optical Gravitational Lensing Experiment. Although the overall incidence rate is very low, about $1$ per cent in each of the Magellanic Clouds, in the case of the SMC and pulsation periods between $12$ and $16$\thinspace d the incidence rate is nearly $40$\thinspace per cent. On the other hand, in the LMC the highest incidence rate is $5$\thinspace per cent for pulsation periods between $8$ and $14$\thinspace d, and the overall amplitude of the effect is smaller. It indicates that the phenomenon is metallicity dependent. Typical modulation periods are between $70$ and $300$\thinspace d. In nearly all stars the mean brightness is modulated, which, in principle, may influence the use of classical Cepheids for distance determination. Fortunately, the modulation of mean brightness does not exceed $0.01$\thinspace mag in all but one star. Also, the effect averages out in typical observations spanning a long time base. Consequently, the effect of modulation on the determination of the distance moduli is negligible. The relative modulation amplitude of the fundamental mode is also low and, with one exception, it does not exceed $6$\thinspace per cent. The origin of the modulation is unknown. We draw a hypothesis that the modulation is caused by the 2:1 resonance between the fundamental mode and the second overtone that shapes the famous Hertzsprung bump progression.
\end{abstract}


%%%%%%%%%%%%%%%%%%%%%%%%%%%%%%%%%%%%%%%%%%%%%%%%%%

%%%%%%%%%%%%%%%%% BODY OF PAPER %%%%%%%%%%%%%%%%%%

\section{Introduction}\label{sect:intro}
%%%%%%%%%%%%%%%%%%%%%%%%%%%%%%%%%%%%%%%%

Classical Cepheids are among the most important variable stars. As primary distance indicators, they are of key importance for stellar astrophysics, galactic astrophysics and cosmology. Because of the very characteristic light curve shape, large amplitude variation and high intrinsic brightness, long-period fundamental mode pulsators are the most useful standard candles. They are also believed to be very regular pulsators, without any large-amplitude modulation of pulsation or additional non-radial modes excited. 

To the contrary, first overtone Cepheids are less stable. Irregular and often fast changes of pulsation period are more common among first overtone than among fundamental mode stars \citep[e.g.][]{poleski}. Additional periodicities, with periods in the $(0.6,\,0.65)P_1$ range, are frequently detected in these stars \citep[e.g.][]{mk09,o3_smc_cep,o4_mc_cep_pec,ss16} and, according to model proposed by \cite{wd16}, are interpreted as due to non-radial modes of moderate degrees. \cite{mk09} reported modulation in several double-overtone Cepheids (simultaneously pulsating in the first and second radial overtones) of the Magellanic Clouds. Recently, \cite{o4_mc_cep_pec} also reported modulation of pulsation in a few classical Cepheids in the Magellanic Clouds. These are either double-overtone pulsators or single-periodic, first overtone pulsators. In addition, V473~Lyr is a very peculiar variable, identified as a second overtone classical Cepheid with modulation \citep[for the recent analysis see][]{ms14}. 

Precise space photometry indicates, that tiny cycle-to-cycle irregularities might occur in classical Cepheids of both modes. \cite{derekas12} detected cycle-to-cycle fluctuations in the pulsation period (of up to $0.02$\thinspace d) in V1154~Cyg -- the only genuine Cepheid in the original {\it Kepler} field \citep{szabo11}. Light curve stability was studied in two classical Cepheids observed with {\it MOST} \citep{evans}. Cycle-to-cycle light curve variation, more pronounced for first overtone pulsator was detected. V1154~Cyg was revisited by \cite{kanev} and \cite{derekas17} who reported the periodic ($\approx\!159$\thinspace d) modulation of the light curve shape. The effect is tiny -- amplitude of the highest modulation peak in the frequency spectrum reported by \cite{derekas17} is $0.6$\thinspace mmag. Long-period modulation is also suspected in fundamental mode Cepheid T~Vul, which was observed with {\it BRITE} Constellation \citep{s17BRITE}. 

Radial velocity measurements also suggest possible modulation in the pulsation of classical Cepheids. \cite{anderson14} studied the radial velocities for two first overtone Cepheids and detected smooth variation of the amplitude and shape of the radial velocity curves on a time scale of years. Data were not sufficient to conclude whether the modulation is periodic. To the contrary, two fundamental mode Cepheids studied by \cite{anderson14} show cycle-to-cycle variations in their radial velocity curves and variation of spectral line features \citep{anderson16}. 

We note that modulation is common in other group of classical pulsators -- in RR~Lyrae stars, in which it is known as the Blazhko effect. Space photometry and top-quality ground-based data indicate that the incidence rate of the Blazhko effect in fundamental mode RR~Lyr stars (RRab stars) is as high as $50$\thinspace per cent \citep[for a review see][]{geza}.

In this paper we report the discovery of periodic modulation in more than 50 fundamental mode classical Cepheids observed by the Optical Gravitational Lensing Experiment \citep[OGLE, e.g.][]{o4} in the Small Magellanic Cloud (SMC) and in the Large Magellanic Cloud (LMC). The effect is much more pronounced than any other instability reported in the light curves of fundamental mode Cepheids so far. The resulting changes in the mean brightness may be as large as $0.01$\thinspace mag in the extreme cases.

\section{Data analysis}
%%%%%%%%%%%%%%%%%%%%%%%

We analyse the publicly available data for the fundamental mode classical Cepheids of the OGLE Magellanic Cloud collection \citep{o3_lmc_cep,o3_smc_cep,o4_mc_cep}. The collection counts 2739 Cepheids in the SMC and 2429 Cepheids in the LMC. We analyse the $I$-band data only which are much more numerous than the $V$-band data. The OGLE-III and OGLE-IV data are provided in the separate files and we analyse them separately. We note that the OGLE-III data file also contains the data from the second phase of the OGLE. 

Data analysis proceeds in two steps. In the first step, we select the stars in which modulation is suspected using the semi-automatic analysis. For each star, 14th order Fourier series is fitted to the data. Severe outliers ($>\!6\sigma$) are removed. Second order polynomial is fitted to the residual data and subtracted from the original data. This way, we remove or reduce the possible slow trends from the data. Fourier series is fitted to the detrended and cleaned data again and Fourier transform of the residuals is stored in the graphical form for inspection. Featureless spectra, i.e. those in which the $\sn\!<\!4.0$ for the highest detected peak, are marked as uninteresting and not inspected. In addition, the data are phased and light curve with Fourier fit is stored in the graphical form.

The number of Cepheids in which $\sn\!>\!4.0$ for the highest peak in the residual spectrum is still rather high. Such peaks still may happen by chance, at random locations in the spectrum. In the long-baseline OGLE data we consider here, trends are also quite common, and sometimes the second-order polynomial is insufficient to remove them. Peaks at close-to-integer frequencies are also common artefact of the ground-based data. These give rise to power excesses, or clumps of peaks, present in the very low frequency range and at close-to-integer frequencies -- all mutual daily aliases. The majority of these cases are easy to identify by inspection of the stored frequency spectrum plots.

\begin{figure}
\includegraphics[width=\columnwidth]{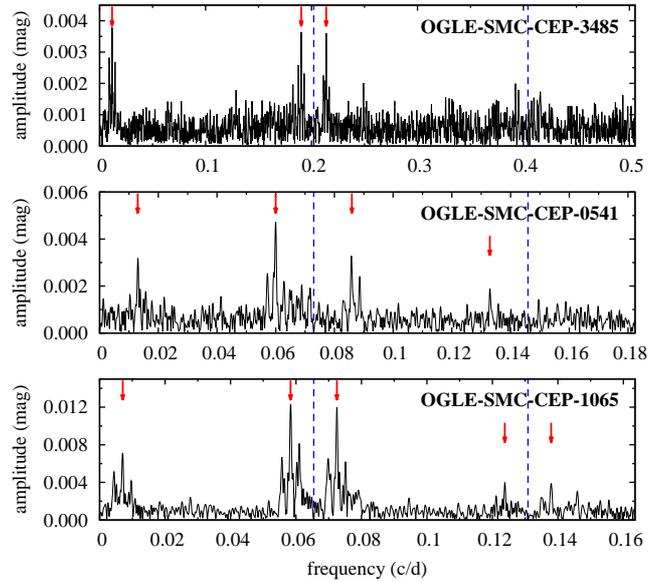}
\caption{Frequency spectra for three modulated Cepheids from the SMC after prewhitening with the fundamental mode and its harmonics (dashed lines). Modulation side peaks are marked with arrows.}
\label{fig:fspSMC}
\end{figure}

\begin{figure}
\includegraphics[width=\columnwidth]{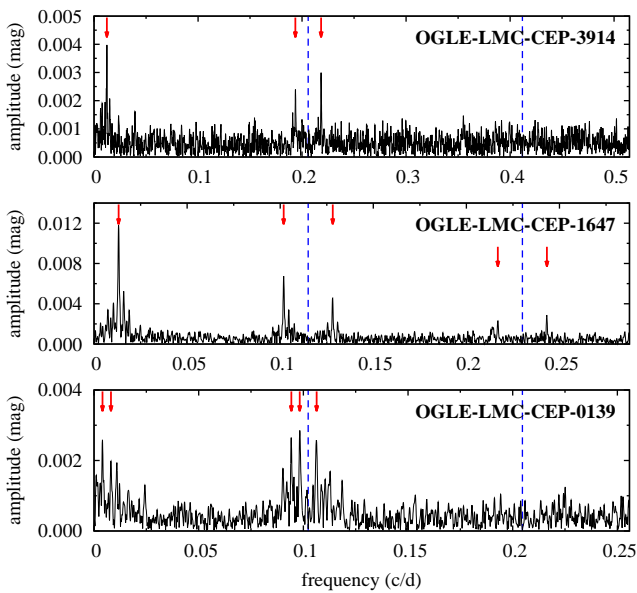}
\caption{The same as Fig.~\ref{fig:fspSMC}, but for three modulated stars from the LMC.}
\label{fig:fspLMC}
\end{figure}

Among the remaining cases, those illustrated in Figs.~\ref{fig:fspSMC} and \ref{fig:fspLMC} are the most interesting. In the frequency spectrum we detect a few significant peaks, frequencies of which are not independent: in the low frequency range, symmetrically placed around fundamental mode frequency, $\fF$, around $2\fF$ and possibly around higher order harmonics. Such peaks may result either because of genuine modulation of pulsation or because of excitation of non-radial modes. The observed effect is the same -- light curve is periodically modulated. Later on (Sect.~\ref{ssect:modulatedQM}), we argue that genuine modulation is the most plausible explanation. Hence, the detected phenomenon is called modulation in the following. In this case, frequencies of the detected peaks are denoted as: $\fm$, for the peak at the low frequency range, $\fF\pm\fm$, for the peaks at the fundamental mode frequency, $2\fF\pm\fm$, for the peaks at the harmonic, etc. Modulation period is $\Pm=1/\fm$.

All suspected cases are carefully analysed on a case-by-case basis, following a standard consecutive prewhitening technique. Details may be found in our previous publications  e.g. in \cite{rs15a}. For a reference, the data are fitted with the sine series of the following form:
\begin{equation}
m(t)=A_0+\sum_{i}A_i\sin(2\uppi\nu_it+\phi_i)\,,\label{eq}
\end{equation}
using non-linear least-square procedure. In the above series, we include all frequencies identified with the help of Fourier transform in the consecutive prewhitening steps: fundamental mode and its harmonics, $k\fF$, modulation side peaks as given above, i.e. $\fm$, $k\fF\pm\fm$ and possibly higher order components of the multiplets. Frequencies, amplitudes and phases are all adjusted. We stress that our approach is conservative. We consider the star as modulated only if at least two peaks that can be associated with modulation of the same period are detected in the spectrum with $\sn\!>\!4$. In addition, in each case we transform the magnitude data to fluxes to check whether these peaks are intrinsic to the star and are not due to contamination. In case additional signal is due to contamination, the peaks at linear combination frequencies (e.g. at $k\fF\pm\fm$) are not present in the flux data\footnote{OGLE-SMC-CEP-1526, OGLE-SMC-CEP-2893 and OGLE-SMC-CEP-3115 are excellent examples. Fourier analysis indicates, that these stars are beat pulsators,  pulsating in two frequencies simultaneously. A few strong combination frequencies are present in the data, but these disappear as soon as magnitude data are transformed into fluxes. Consequently, the three stars are blends of two high-amplitude variables, each.}.

\begin{figure}
\includegraphics[width=\columnwidth]{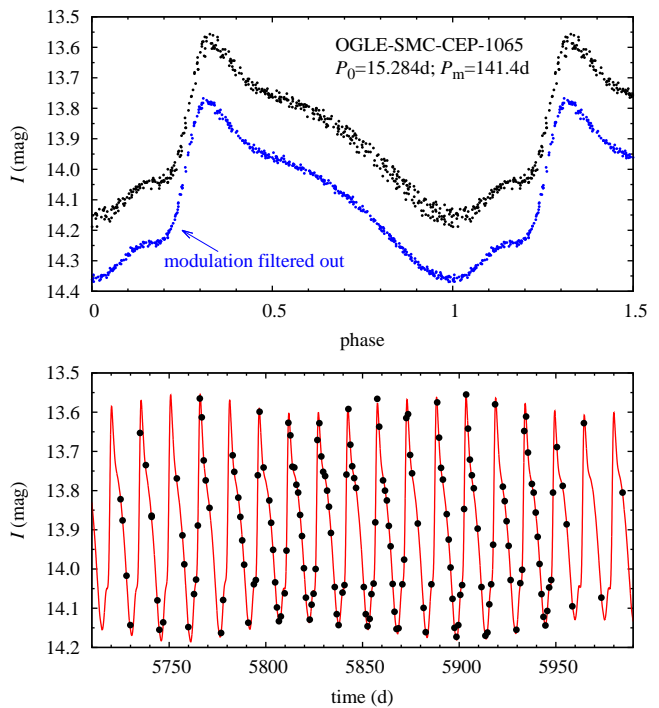}
\caption{Light curve for OGLE-SMC-CEP-1065. Phased light curve is plotted in the top panel (black points). For comparison, we also plot the light curve with modulation filtered out (blue points, shifted by $0.2$\thinspace mag). In the bottom panel we show the section of the light curve with fitted model overplotted (solid line). Time is expressed as $t={\rm HJD}-245\,0000\,{\rm d}$. }
\label{fig:lcSMC}
\end{figure}

\begin{figure}
\includegraphics[width=\columnwidth]{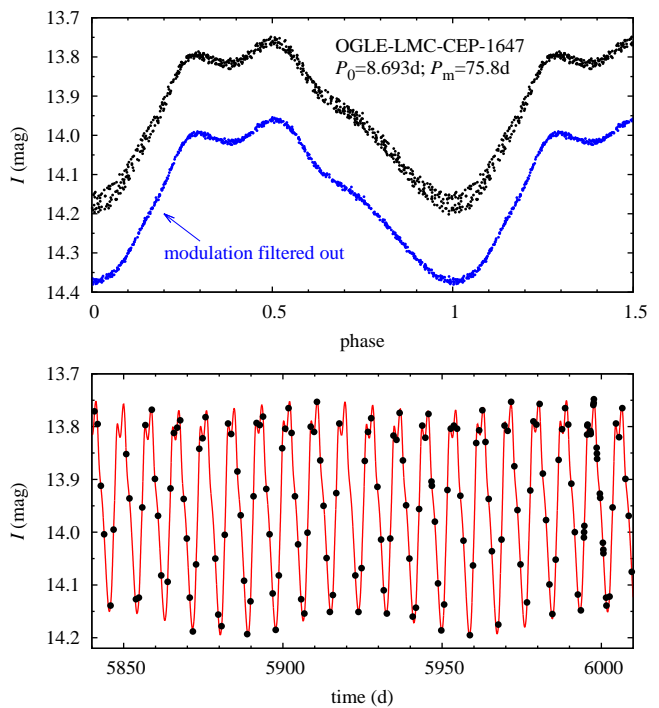}
\caption{The same as Fig.~\ref{fig:lcSMC} but for OGLE-LMC-CEP-1647.}
\label{fig:lcLMC}
\end{figure}

 In the majority of cases, the modulation is weak and the additional signals are in the mmag range. Only in a few cases the signature of modulation is directly visible in the light curve. The strongest cases from the SMC and LMC are illustrated in Figs.~\ref{fig:lcSMC} and \ref{fig:lcLMC}, respectively. The top panels show the phased light curves: raw and with modulation filtered out (by subtracting from the data the sine series with all but $k\fF$-terms, see eq.~\ref{eq}). The bottom panels show a section of the photometric time series with full model fitted to the data. 

\begin{figure}
\includegraphics[width=\columnwidth]{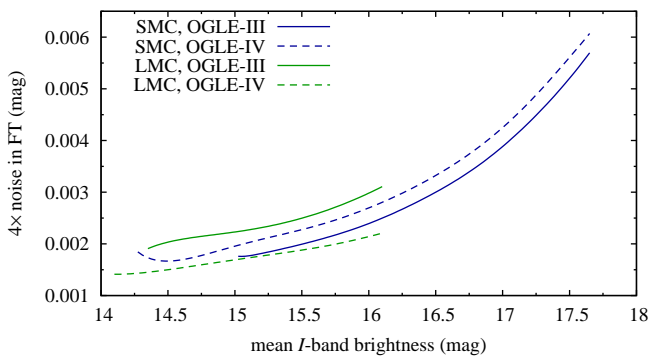}
\caption{Detection limit in the OGLE SMC and LMC data as a function of mean brightness of the Cepheids. See text for details.}
\label{fig:noise}
\end{figure}

Before we discuss the results, we first estimate the detection limit in the OGLE data in a way similar to our previous work for the first overtone Cepheids in the SMC \citep{ss16}. Estimates are done independently for the LMC and SMC and for the OGLE-III and OGLE-IV data. For all fundamental mode Cepheids in a given sample, the photometric data are prewhitened with a 14th order Fourier series. Fourier transform of the residuals is computed and average signal in the $(0,\, 6\fF)$ range -- the noise, ${\rm N}$, is computed. Data on spectra in which no strong signal remains ($\sn\!<\!5$) are used to construct the four-times-noise (the detection limit) versus the mean $I$-band brightness plots. The resulting smoothed running averages are plotted in Fig.~\ref{fig:noise} for the four samples. As expected, the noise increases as mean brightness (and pulsation period; $P-L$ relation) decreases. Differences between the four considered data samples result from different observing strategy (different data sampling) for the SMC and for the LMC during the two phases of the OGLE project. For the LMC, the noise level is lower in the OGLE-IV data and consequently, this data set allows us to detect weaker signals. At the luminosity range covered by the LMC Cepheids, the typical detection limit is $1.5\!-\!2$\thinspace mmag. In the case of the SMC, the difference between the OGLE-III and OGLE-IV data is not as pronounced and is in the opposite direction -- the noise level is typically lower in the OGLE-III data. The SMC Cepheids are fainter; the detection limit vary from $2$ to $5$\thinspace mmag. In practice, for all interesting stars, the OGLE-III and OGLE-IV data were analysed independently. Comparison of the results, yields information on the stability of the detected phenomena. Occasionally, when long-period modulation was suspected, the OGLE-III and OGLE-IV data were merged. 

\section{Results}\label{sec:results}
%%%%%%%%%%%%%%%%%%%%%%%%%%%%%%%%%%%%

Data on the identified modulated stars from the SMC and from the LMC are collected in Tabs.~\ref{tab:smc} and \ref{tab:lmc}, respectively. First five columns contain: star's id, indication of data used in the analysis, mean $I$-band brightness as given in the OGLE catalog, the pulsation and the modulation periods. Sixth column lists the Fourier amplitude of the fundamental mode, and consecutive five columns list the amplitudes of modulation peaks detected at $\fm$, $\fF-\fm$, $\fF+\fm$, $2\fF-\fm$ and $2\fF+\fm$, respectively. The largest amplitude among the above modulation peaks is indicated with the bold face font. Asterisk indicates that the associated variability is non-coherent -- significant and unresolved power remains after the prewhitening. If other modulation peaks were detected, then they are listed in the last column of Tabs.~\ref{tab:smc} and \ref{tab:lmc} (notes on frequency content, denoted with capital letters), which also contains remarks on individual stars, as listed in the tables' caption. In case secondary modulation was detected, respective data are given in additional row. 

Numerical results collected in the tables result from the analysis of data indicated in the second column, in which `o3'/`o4' stand for the OGLE-III and OGLE-IV data, respectively. For the LMC, our primary choice is OGLE-IV, as these data, on average, assure a lower noise level. As discussed in the previous section, for the SMC, the OGLE-III and OGLE-IV data are of comparable quality, OGLE-III data being slightly better. For the SMC, OGLE-III data are used for stars with longer modulation periods. Also, if the number of detected modulation peaks is different in the two data sets, we list the results for the data in which more side peaks were detected. We stress that in all cases we checked both data sets. `o3C'/`o4C' in the remarks column of Tabs.~\ref{tab:smc} and \ref{tab:lmc} indicate, that modulation (with the same period) was confirmed also in the other data set, although sometimes the detection was weak (e.g. only one side peak that could be associated with the modulation was detected). This is indicated with a colon. A comparison of the results for the two data sets indicates, that modulation is a stable phenomenon; in the majority of cases, modulation period and amplitudes of the side peaks agree within formal errors. `o3X'/`o4X' indicate that modulation was not detected in the other data set. In all such cases we have checked that it was a consequence of a too large noise level. `o4-nr' indicates that modulation is not resolved in the OGLE-IV data and `\cancel{o3}'/`\cancel{o4}' indicate that data were not collected during the respective phase of the OGLE.

In some stars we also detected additional significant periodicities. This is marked with `fX' or `sm' in the remarks column. In the latter case, the additional power was detected close to the radial mode frequency, indicating a possible secondary modulation. In no case however, we detect additional peaks that would correspond to linear frequency combination with the previously detected peaks, in particular with the radial mode frequency. In the majority of cases, these detections are also weak, with $\sn<5$. Consequently, these peaks might appear in the frequency spectrum by chance, or may result from contamination. They are not discussed further.

Data in Tabs.~\ref{tab:smc} and \ref{tab:lmc} are intentionally sorted by the increasing pulsation period -- to visualize the change in amplitudes of the modulation peaks as period changes. We also note that in the case of LMC, only in 8 cases we detect modulation peaks close to $2\fF$. Still we decided to leave the form of Tab.~\ref{tab:lmc} the same as for Tab.~\ref{tab:smc} to explicitly visualize this difference between the modulated stars of the SMC and LMC.

\begin{table*}
\centering
\caption{Basic data on modulated Cepheids detected in the SMC. Consecutive columns contain: star's id (last four digits of the full OGLE-SMC-CEP-nnnn id), indication of data used in the analysis, mean $I$-band brightness from the OGLE catalog, pulsation period, modulation period and amplitudes of the peaks detected at $\fF$, $\fm$, $\fF-\fm$, $\fF+\fm$, $2\fF-\fm$ and $2\fF+\fm$. In the last column remarks and notes on individual stars and their frequency content are given. Remarks are: `o3C'/`o4C' -- modulation confirmed in the OGLE-III/OGLE-IV data; `o3X'/`o4X' --  modulation not confirmed in the OGLE-III/OGLE-IV data, because of their lower quality; `o4-nr' -- modulation not resolved in the OGLE-IV data; `\cancel{o3}'/`\cancel{o4}' -- no OGLE-III/IV data; `fX' -- additional significant periodicity present in the spectrum (but without combination frequencies); `sm' -- as `fX' but close to the radial mode frequency, indicating a possible secondary modulation. Notes on individual stars are: 1 -- incomplete phase coverage due to near-to-integer fundamental mode period; 2 -- light curve with very low amplitude; 3 -- incomplete phase coverage. Notes on frequency content are: A, $3\fF\!+\!\fm$, $4\fF\!+\!\fm$; B, $3\fF\!-\!\fm$; C, $\fm/2$; D, $2\nu_{\rm m2}$; E, $2\fF\!+\!2\fm$, $3\fF\!+\!2\fm$; F, complete septuplets detected at $\fF$ and $2\fF$; G, $\fF\!+\!\nu_{\rm m1}\!-\!\nu_{\rm m2}$, $\fF\!-\!2\nu_{\rm m1}\!+\!\nu_{\rm m2}$; H, $2\fm$.}
\label{tab:smc}
%fcOK - final check OK!
\begin{tabular}{lllr@{.}lr@{}lrrrrrrl}
   &      &  &\multicolumn{2}{c}{}  & \multicolumn{2}{c}{}    & \multicolumn{6}{c}{amplitudes of peaks at: (mag)} & \\
id & data & $I$\thinspace(mag) &\multicolumn{2}{c}{$\Pf$ (d)} & \multicolumn{2}{c}{$\Pm$ (d)} & $\fF$ & $\fm$ &  $\fF\!-\!\fm$ &   $\fF\!+\!\fm$ &   $2\fF\!-\!\fm$ & $2\fF\!+\!\fm$ & remarks \\ 
\hline
\smccep3267 & o4 & 15.410 &  3&693335(6) &   76&.9(2)   & 0.1895 &      0.0022  & {\bf 0.0023} & 0.0019 &  -     & - & o3C: \\ 
\smccep4220 & o4 & 15.489 &  3&98808(3)  &   65&.4(1)   & 0.1481 & {\bf 0.0071} &      0.0058  & 0.0054 &  -     & - & o3C, 1 \\
\smccep3379 & o4 & 15.242 &  4&363804(7) &   88&.3(2)   & 0.2076 &      0.0036  & {\bf 0.0037} & 0.0027 & 0.0019 & - & o3C \\ 
\smccep2158 & o3 & 15.395 &  4&735925(3) &   71&.82(7)  & 0.2207 &      0.0021  & {\bf 0.0023} & 0.0019 & 0.0013 & - & \cancel{o4}\\ 
\smccep3485 & o4 & 14.876 &  4&947663(8) &   85&.0(1)   & 0.2303 & {\bf 0.0044} & 0.0042 & 0.0039 & 0.0025 & 0.0022 & o3C, A \\ 
\smccep4014 & o4 & 15.110 &  5&46143(4)  &   71&.83(7)  & 0.0810 & {\bf 0.0059} & 0.0040 & 0.0030 &  -     &  -     & o3C, fX, 2 \\ 
\smccep1493 & o3 & 15.240 &  5&954754(7) &   72&.49(3)  & 0.1565 & {\bf 0.0057} & 0.0031 & 0.0025 & 0.0016 &  -     & o4C, fX \\ 
\smccep1746 & o3 & 14.796 &  6&921823(7) &   83&.89(2)  & 0.1505 & {\bf 0.0090} & 0.0053 & 0.0034 & 0.0020 & 0.0015 & o4C,\\ 
\smccep3934 & o4 & 14.485 &  7&16464(4)  &  145&.1(5)   & 0.1297 & {\bf 0.0037} & 0.0021 & 0.0021 & 0.0019 &  -     & o3C \\ 
\smccep0896 & o3 & 15.183 &  7&332460(8) &   27&.770(4) & 0.1939 &      0.0014  & {\bf 0.0068} & 0.0021 & 0.0020 & - & o4C\\ 
\smccep0895 & o4 & 14.457 & 10&44517(4)  &   99&.8(2)   & 0.2007 &      0.0021  & {\bf 0.0042} & - & 0.0019 & - & o3C:, B \\ 
\smccep0705 & o4 & 14.588 & 10&48503(6)  &  137&.5(4)   & 0.1933 & {\bf 0.0060} &      0.0019  & - &  -     & - & o3C, C, 3\\  
\smccep1051 & o3 & 14.414 & 11&77064(3)  &  351&(2)     & 0.1541 &       -      & {\bf 0.0032} &      0.0025  & - & - & o4C \\ 
\smccep0387 & o4 & 13.982 & 13&0909(1)   &   32&.98(3)  &*0.2018 &       -      &   -    & {\bf 0.0053} & - &  0.0020 & o3C, fX \\ 
   ~        & o4 & 13.982 & 13&0909(1)   &  115&.3(4)   &*0.2018 &       -      & 0.0023 & {\bf 0.0025} & - &  0.0019 & o3C, D  \\ 
\smccep4478 & o3 & 13.617 & 13&19049(5)  &  317&.5(9)   & 0.1840 &      0.0043  & {\bf 0.0069} & 0.0061  & - & - & o4C \\ 
\smccep1247 & o3 & 14.335 & 13&30134(2)  & 1139&(3)     &*0.1976 &      0.0077  & 0.0100 & {\bf 0.0104} & 0.0035 & 0.0033 &  o4-nr\\ 
\smccep4393 & o4 & 13.801 & 13&4635(1)   &  182&.7(5)   &*0.1958 &      0.0037  & {\bf 0.0064} &      0.0056  & - & - & o3C\\  
\smccep0368 & o3 & 13.993 & 13&47823(2)  &  133&.0(1)   & 0.1921 &      0.0018  & {\bf 0.0030} &  -  & 0.0010 & - & o4C, B\\ 
\smccep0541 & o4 & 14.024 & 13&70563(7)  &   77&.1(1)   & 0.2369 &      0.0034  & {\bf 0.0051} & 0.0033 & 0.0019 &  -     & o3C \\ 
\smccep3490 & o4 & 13.601 & 13&78652(7)  &  159&.9(3)   & 0.2135 &      0.0015  & - &*{\bf 0.0080} & - & 0.0020 & o3X, E\\ 
\smccep2031 & o3 & 13.765 & 14&57265(2)  &  143&.3(3)   &*0.2219 &       -      &       -      & {\bf 0.0029}  &  -     & 0.0018 & o4C:\\ 
\smccep2574 & o3 & 13.808 & 14&59147(5)  & 1287&(6)     & 0.1887 &      0.0045  &      0.0095  & {\bf 0.0098}  & 0.0029 & 0.0026 & o4-nr, F\\ 
\smccep4906 & o4 & 13.181 & 15&1683(2)   &  198&(1)     & 0.2144 &      0.0038  &       -      & {\bf 0.0062}  &  -     &  -     & \cancel{o3}, sm \\ 
\smccep1065 & o4 & 13.897 & 15&2840(1)   &  141&.4(1)   & 0.2316 &      0.0068  &      0.0121  & {\bf 0.0122}  & 0.0032 & 0.0043 & o3C, G\\ 
   ~        & o4 & 13.897 & 15&2840(1)   &   82&.4(2)   & 0.2316 &       -      &       -      &*{\bf 0.0036}  &  -     & 0.0022 & \\
\smccep1569 & o3 & 13.703 & 15&64815(2)  & 1345&(5)     & 0.2289 &      0.0064  & {\bf 0.0088} &      0.0085   & 0.0043 & 0.0037 & o4-nr, H \\ 
\smccep1385 & o3 & 13.841 & 15&82210(5)  &  162&.6(3)   & 0.2480 &      0.0039  &       -      &*{\bf 0.0047}  & -      & 0.0030 & o4C\\ 
\smccep0152 & o4 & 13.401 & 15&8375(1)   &  132&.9(3)   & 0.1889 &       -      &       -      & {\bf 0.0055}  & -      & 0.0019 & o3C\\ 
\smccep4697 & o4 & 13.088 & 18&9014(2)   &  190&.9(1.5) & 0.2119 &      0.0025  &       -      & {\bf 0.0042}  & -      &  -     & \cancel{o3}\\ 
\smccep1107 & o4 & 13.646 & 24&4839(2)   &  154&.8(3)   & 0.2621 & {\bf 0.0089} & 0.0067 &  0.0047 & 0.0031 & - & o3C, sm \\
 \hline  
\end{tabular}
\end{table*}

\begin{table*}
\centering
\caption{The same as Tab.~\ref{tab:smc}, but for the modulated Cepheids in the LMC. Notes on individual stars are: 1 -- OGLE-III light curve corrupted with many dim outliers. Notes on frequency content are: A, $2\fm$; B, $\nu_{\rm m2}-\nu_{\rm m1}$, $2\nu_{\rm m2}-\nu_{\rm m1}$; C, $2\fm$; D, $3\fF+\fm$, $3\fF-\fm$; E, quintuplet at $\fF$, $2\fm$, $\fm/2$; F, $\fF-3\fm$, $\fF-2\fm$, $\fF+2\fm$, $2\fm$, $3\fm$, $6\fm$.}
\label{tab:lmc}
\begin{tabular}{lllr@{.}lr@{}lrrrrrrl}
   &      &  &\multicolumn{2}{c}{}   & \multicolumn{2}{c}{}  & \multicolumn{6}{c}{amplitudes of peaks at: (mag)} & \\
id & data & $I$\thinspace (mag) &\multicolumn{2}{c}{$\Pf$ (d)} & \multicolumn{2}{c}{$\Pm$ (d)} & $\fF$ & $\fm$ &  $\fF\!-\!\fm$ &   $\fF\!+\!\fm$ &   $2\fF\!-\!\fm$ & $2\fF\!+\!\fm$ & remarks \\ 
\hline
\lmccep3003 & o4 & 15.446 &  2&626009(2) &  84&.2(2) & 0.1688 & 0.0016 & {\bf 0.0020} & 0.0019 & - & 0.0014 & o3X \\ 
\lmccep1800 & o4 & 15.175 &  3&066685(2) &  71&.6(1) & 0.2264 & 0.0020 & {\bf 0.0020} & 0.0020 & - & - & o3C:\\ 
\lmccep0723 & o4 & 15.219 &  3&886986(6) &  95&.2(2) & 0.1572 & 0.0020 & {\bf 0.0023} & 0.0016 & - & - & o3C:  \\
\lmccep0852 & o4 & 14.907 &  4&334864(8) & 110&.7(4) & 0.2044 & {\bf 0.0027} & 0.0021 & 0.0016 & - & - & o3C \\ 
\lmccep3914 & o4 & 14.810 &  4&85627(1)  &  80&.9(1) & 0.1493 & {\bf 0.0041} & 0.0023 & 0.0027 & 0.0017 & 0.0017 & \cancel{o3}, fX \\
\lmccep1888 & o4 & 14.578 &  5&407258(9) & 126&.8(7) & 0.1854 & {\bf 0.0013} & 0.0012 & -      & - & - & o3X \\
\lmccep2635 & o4 & 14.481 &  6&27203(2)  & 131&.3(4) & 0.1311 & {\bf 0.0028} & 0.0010 & 0.0014 & - & - & o3C \\ 
\lmccep2405 & o4 & 15.106 &  6&53185(3)  & 118&.0(3) & 0.0938 & {\bf 0.0028} & 0.0014 & 0.0015 & - & - & o3C\\ 
\lmccep2866 & o4 & 14.587 &  6&61329(2)  &  97&.2(2) & 0.1461 & - & {\bf 0.0018} & 0.0016 & - & - & o3C:, sm\\ 
\lmccep2008 & o4 & 14.184 &  7&51896(4)  & 124&.5(3) & 0.0951 & {\bf 0.0038} & 0.0016 & 0.0017 & - & - & o3C, B\\ 
    ~       & o4 & 14.184 &  7&51896(4)  &  83&.7(2) & 0.0951 & {\bf 0.0021} & 0.0010 & - & - & - & o3C\\
\lmccep2775 & o4 & 14.165 &  7&96240(3)  &  62&.4(1) & 0.1556 & {\bf 0.0016} & - & 0.0012 & - & - & o3X, C \\
\lmccep4176 & o4 & 13.936 &  8&33828(3)  &  86&.3(2) & 0.1764 & -            & {\bf 0.0028} & - & 0.0013 & - & \cancel{o3}, fX\\ 
\lmccep1647 & o4 & 13.939 &  8&69305(2)  &  75&.84(2) & 0.1906 & {\bf 0.0126} & 0.0069 & 0.0046 & 0.0024 & 0.0025 & o3C, D\\
\lmccep1956 & o4 & 14.280 &  8&80145(4) & 130&.8(2)  & 0.1476 & {\bf 0.0051} & 0.0033 & 0.0023 & 0.0010 & - & o3C\\ 
\lmccep1083 & o4 & 13.913 &  8&89967(2) & 264&(1)    & 0.2016 & 0.0013 & {\bf 0.0017} & 0.0014 & - & - & o3X, E\\ 
\lmccep0139 & o4 & 14.142 &  9&77362(5) & 247&.5(5)  & 0.1820 & 0.0026 & {\bf 0.0029} & 0.0024 & - & - & o3C, F\\ 
\lmccep3574 & o4 & 13.856 & 11&3219(4)  &  48&.16(5) & 0.0321 & - & {\bf 0.0041} & 0.0028 & - & - & \cancel{o3}, fX  \\ 
\lmccep0561 & o4 & 13.547 & 11&67042(4) &  94&.3(2)   & 0.2214 & 0.0016 & - &*{\bf 0.0040} & - & 0.0014 & o3C:, sm \\ 
\lmccep0356 & o4 & 13.694 & 12&68416(4) & 153&.7(7)  &*0.2656 & 0.0017 & - & {\bf 0.0030} & - & 0.0016 & o3C\\ 
\lmccep1539 & o4 & 13.504 & 13&51319(7) & 195&.4(6)  & 0.2100 & {\bf 0.0056} & 0.0025 & - & - & - & o3C \\ 
\lmccep2023 & o4 & 13.001 & 16&2256(1)  &  75&.0(2)  & 0.2596 & -      & - & {\bf 0.0063} & - & 0.0039 & o3X, 1\\ 
\lmccep0935 & o4 & 12.898 & 21&1163(2)  & 213&(2)    & 0.2685 & 0.0022 & {\bf 0.0029} & - & - & - & o3X \\
\hline
\end{tabular}
\end{table*}

\subsection{Are these genuine fundamental mode Cepheids?}
%%%%%%%%%%%%%%%%%%%%%%%%%%%%%%%%%%%%%%%%%%%%%%%%%%%%%%%%%

Before we discuss the properties of the detected modulation, we first check whether the modulated stars are genuine, fundamental mode Cepheids of the Magellanic Clouds. To this aim, we analyse the period-luminosity and colour-magnitude diagrams and investigate the light curve shapes. 

\begin{figure}
\includegraphics[width=\columnwidth]{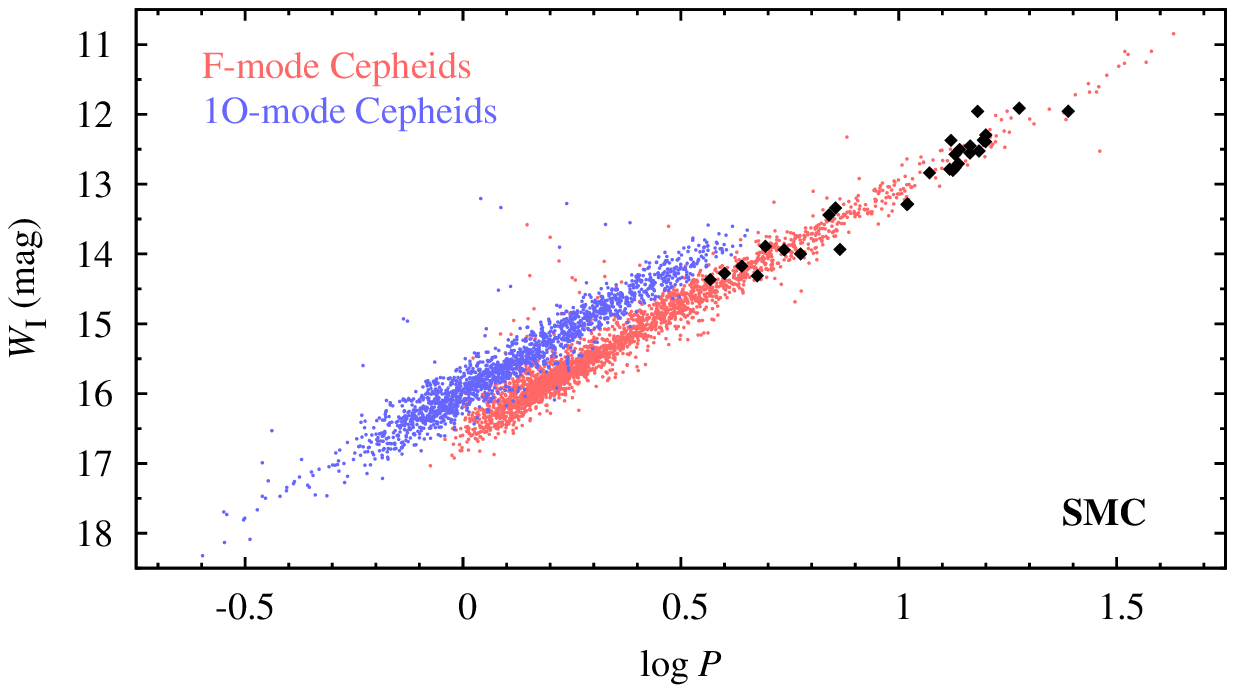}\\
\includegraphics[width=\columnwidth]{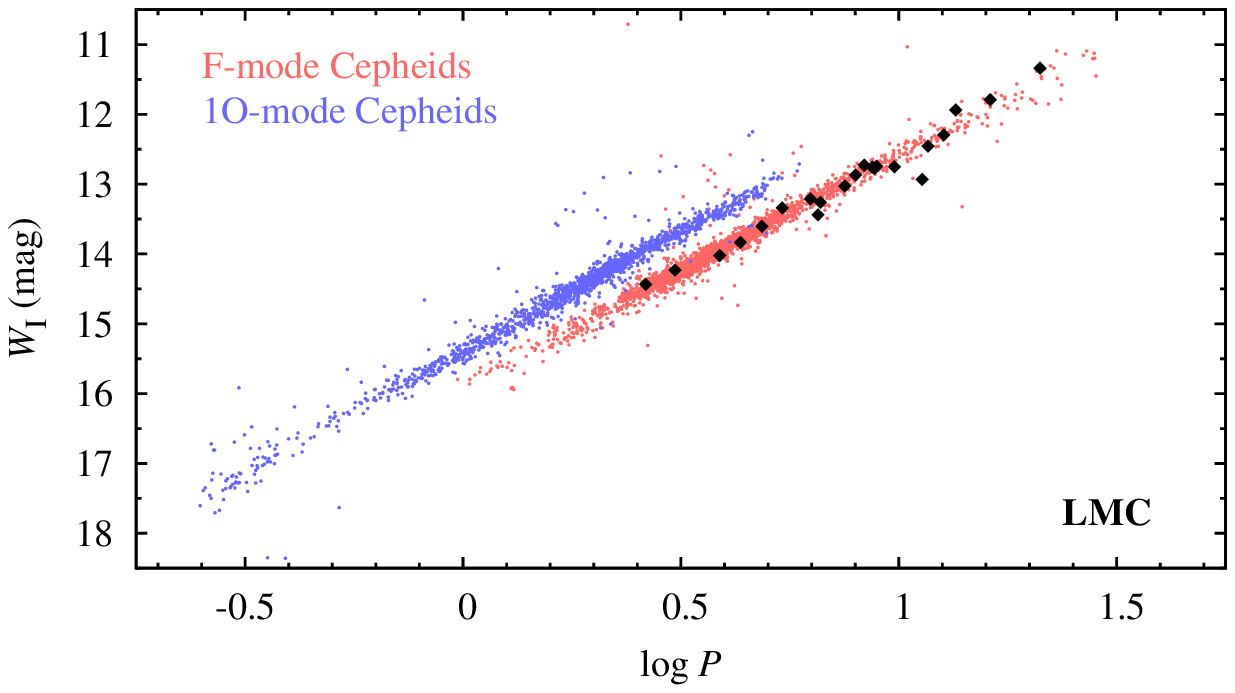}
\caption{Period-luminosity (Wesenheit index) relations for classical Cepheids of the SMC (top panel) and the LMC (bottom panel). Sequences for fundamental mode (reddish dots) and first overtone (bluish dots) Cepheids are plotted. Large diamonds correspond to modulated Cepheids. Mean $I$- and $V$-band brightnesses of the stars were taken from the OGLE catalogue \citep{o4_mc_cep}.}
\label{fig:pl}
\end{figure}

\begin{figure}
\includegraphics[width=\columnwidth]{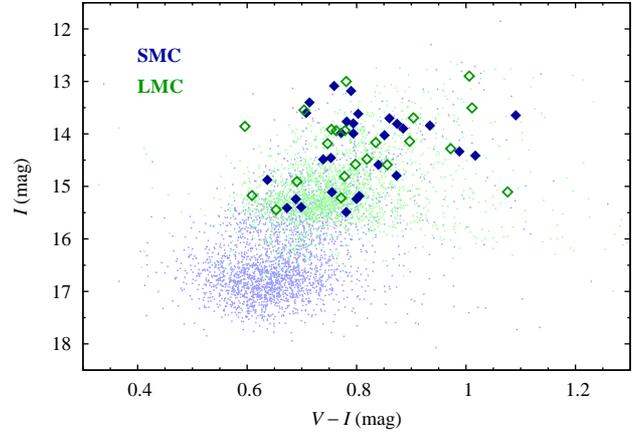}
\caption{The observed $I$ versus $V-I$ colour-magnitude diagram for fundamental mode Cepheids in the SMC (blue) and in LMC (green). Modulated stars are marked with large symbols. Mean $I$- and $V$-band brightnesses of the stars were taken from the OGLE catalogue \citep{o4_mc_cep}.}
\label{fig:cmd}
\end{figure}

The period-luminosity diagrams are plotted in Fig.~\ref{fig:pl} for the SMC (top panel) and for the LMC (bottom panel). On vertical axes, we plot the reddening-free Wesenheit index, $W_I=I-1.55(V-I)$. Small symbols correspond to all Cepheids from the OGLE collection, including the first overtone sample for a reference. In Fig.~\ref{fig:cmd} we show the observed colour-magnitude diagrams for fundamental mode Cepheids of both clouds, marking the modulated stars with large symbols. Based on these two figures we conclude: 

(i) All modulated Cepheids follow the $P-L$ relation defined by the fundamental mode Cepheids of a given Cloud. (ii) All modulated Cepheids, in both Clouds, are brighter than $\approx\!15.5$\thinspace mag. (iii) In the case of the LMC, it means that modulated stars cover nearly the full range of luminosities and periods characteristic for the Cloud, with the exception of faint, short period ($\Pf<2.6$\thinspace d) tail, in which the overall number of fundamental mode Cepheids is small. (iv) In the SMC, the situation is very different. Its population of fundamental mode Cepheids is significantly shifted towards the shorter periods (and fainter luminosities) as compared to the LMC, while the luminosity/period limit for the modulated stars is roughly the same as in the case of LMC. Consequently, modulated stars are not present at the low-luminosity, short-period ($\Pf<3.6$\thinspace d) tail of the SMC, in which the gross of the fundamental mode Cepheids reside. This is not an observational selection effect as we will discuss in more detail in Sect.~\ref{ssect:amps}. (v) At a given luminosity level, the modulated stars cover the full colour range characteristic for a given Cloud. 

\begin{figure}
\includegraphics[width=\columnwidth]{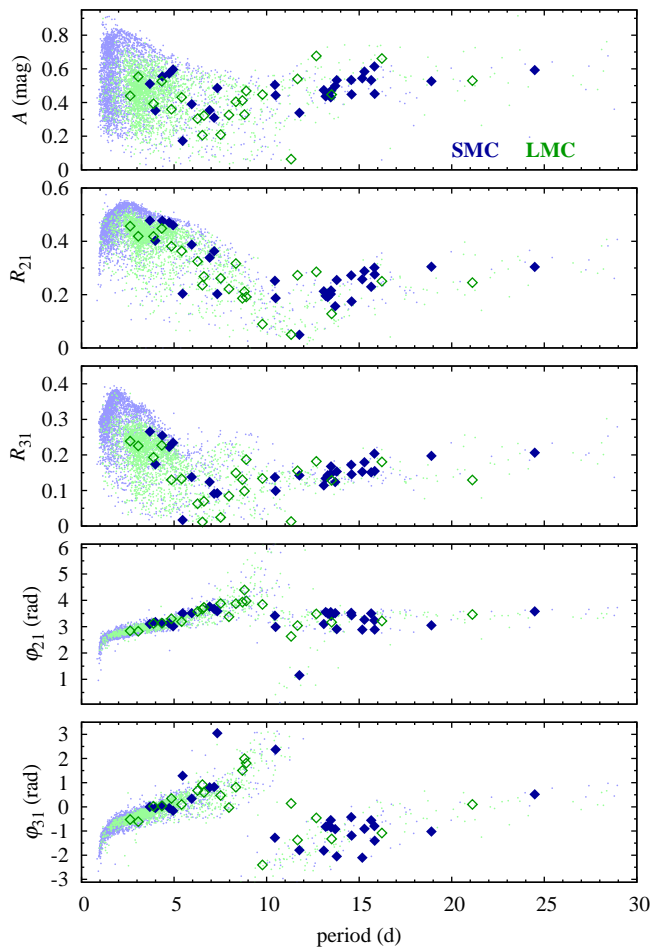}
\caption{Peak-to-peak amplitude (top) and low-order Fourier decomposition parameters of the modulated Cepheids (large symbols) compared with the parameters of all fundamental mode Cepheids from the OGLE collection \citep[small dots,][]{o4_mc_cep}.}
\label{fig:fp}
\end{figure}

In Fig.~\ref{fig:fp} we analyse the light curve shapes for the Cepheids using the low-order Fourier decomposition parameters \citep{sl81}. These are constructed using the amplitudes and phases of the harmonic, $k\fF$-terms in eq.~\eqref{eq}: $R_{k1}=A_k/A_1$, $\varphi_{k1}=\phi_k-k\phi_1$. In Fig.~\ref{fig:fp}, these parameters and peak-to-peak amplitude, $A$, are plotted versus the period for all fundamental mode Cepheids of the Magellanic Clouds, with modulated stars marked with large symbols. Shape of the light curve of classical Cepheids changes with the increasing pulsation period. At the shortest periods, the light curve is essentially saw-tooth variation. At pulsation periods above $\approx\!5$\thinspace d, the bump appears on the descending branch and, as period increases, gradually moves towards ascending branch, distorting the brightness maximum at pulsation periods around $10$\thinspace d. This is the famous Hertzsprung bump progression which is a result of the 2:1 resonance between the fundamental mode and the second overtone, $2\fF=\nu_2$ \citep[e.g.][]{ss76,bk86,bmk90}. Due to this resonant effect, fundamental mode Cepheids follow the characteristic progressions in the Fourier parameter plots in Fig.~\ref{fig:fp}. Modulated stars are no different and follow exactly the same progressions. In the majority of modulated Cepheids the bump feature is clearly visible in the light curve. We note that before computing the Fourier parameters for modulated stars we filtered out the modulation. Modulation, due to its low amplitude, barely affects the light curve shape however, as Figs.~\ref{fig:lcSMC} and \ref{fig:lcLMC} nicely illustrate. In these two stars, characterized by the strongest modulation, bump distorting the phases at maximum light (Fig.~\ref{fig:lcLMC}) and shifted towards the ascending branch (Fig.~\ref{fig:lcSMC}) is obvious. In Sect.~\ref{ssect:modulatedQM} we discuss the possible connection between the modulation and the 2:1 resonance.

The above analysis points that all modulated stars are genuine, fundamental mode Cepheids, properties of which do not differ significantly from the rest of Cepheids in the Magellanic Clouds. Now we will discuss the properties of the detected modulation.

\subsection{Pulsation and modulation periods}
%%%%%%%%%%%%%%%%%%%%%%%%%%%%%%%%%%%%%%%%%%%%%

\begin{figure}
\includegraphics[width=\columnwidth]{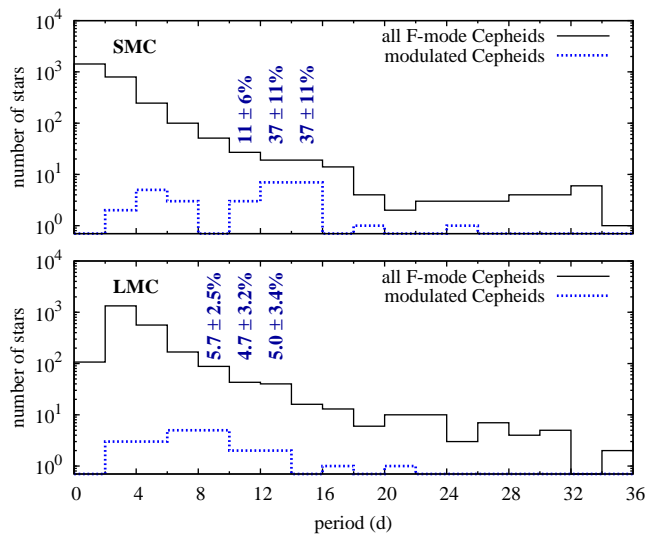}
\caption{Period distribution of all (black solid line) and modulated (blue dotted line) fundamental mode Cepheids in the SMC (top) and in the LMC (bottom). Incidence rate of modulated stars is given only if it exceeds $\approx5$\thinspace per cent in a bin with significant statistics (at least 10 Cepheids). Note the logarithmic scale on vertical axis.}
\label{fig:histo_ic}
\end{figure}

In Fig.~\ref{fig:histo_ic}, we show the pulsation period distribution for all fundamental mode Cepheids (solid lines) and for the modulated ones (dashed lines). Although modulation was detected in only 29 Cepheids of the SMC, which barely exceeds 1 per cent of the sample (top panel of Fig.~\ref{fig:histo_ic}), modulation is very common for pulsation periods in between $12$ and $16$\thinspace d -- nearly $37$\thinspace per cent of the Cepheids show the phenomenon (14 out of 38). Modulation was also found in 10 Cepheids with periods between $2$ and $8$\thinspace d, but the overall number of Cepheids is very large here, so the incidence rate is very low (below $3$\thinspace per cent). No modulations were detected in stars with $7.4\lesssim\Pf\lesssim10.4$\thinspace d and only in two modulated stars the pulsation period is longer than $16$\thinspace d (Tab.~\ref{tab:smc}). Situation is different for the LMC; a glimpse at Tab.~\ref{tab:lmc} shows that the effect is of lower amplitude there (see also below). In no period bin plotted in Fig.~\ref{fig:histo_ic} (bottom panel) the modulation is frequent. The highest incidence rate is at the level of $5\!-\!6$\thinspace per cent for pulsation periods between $8$ and $14$\thinspace d. Only six modulated stars have pulsation periods above $10$\thinspace d. For the remaining 14 modulated stars pulsation periods are shorter. As fundamental mode Cepheids are abundant at shorter pulsation periods, the incidence rate is below $5$\thinspace per cent for all period bins displayed in Fig.~\ref{fig:histo_ic}, in which the number of Cepheids is significant (at least 10 stars).  

\begin{figure}
\includegraphics[width=\columnwidth]{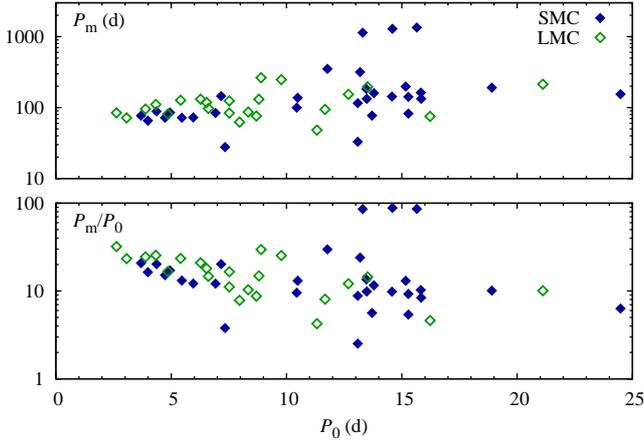}
\caption{Modulation period, $\Pm$ (top panel), and ratio of the modulation to pulsation periods, $\Pm/\Pf$ (bottom panel) versus the pulsation period for Cepheids of the SMC (blue symbols) and the LMC (green symbols). Note the logarithmic scale on vertical axis.}
\label{fig:periods}
\end{figure}

Modulation periods are analysed in Fig.~\ref{fig:periods}. In the top panel we plot the modulation period versus the pulsation period, while ratio of these two periods is plotted in the bottom panel. We observe that there is no difference concerning the modulation periods between the SMC and LMC Cepheids. Modulation periods are typically between $70$ and $300$\thinspace d, $100$\thinspace d seems the most typical value. A slight trend of the increasing modulation period with the increasing pulsation period is well visible. In the bottom panel of Fig.~\ref{fig:periods}, we plot the period ratio, $\Pm/\Pf$. Here, a trend of the decreasing period ratio with the increasing modulation period is present, specifically for pulsation periods below $10$\thinspace d. The modulation period is typically ten times longer than the pulsation period. 

Three SMC stars stand out in Fig.~\ref{fig:periods} -- those with the longest modulation periods, all above $1000$\thinspace d (these are OGLE-SMC-CEP-1247, OGLE-SMC-CEP-2574 and OGLE-SMC-CEP-1569). In fact four additional candidates for long-period modulation were detected in the SMC. These stars are explicitly listed in Tab.~\ref{tab:longp} in which we give the pulsation period, the estimate of the modulation period and additional comments in the last column. The suspected modulation period of $707$\thinspace d in OGLE-SMC-CEP-0359 is well resolved in the data, but the supposed side peaks are mutual 1-yr aliases. After prewhitening with one of the side peaks, a residual power of complex structure remains at the location of the other side peak. Modulation is also resolved in the merged OGLE-III and OGLE-IV data for OGLE-SMC-CEP-0423, leading to modulation period of more than $2800$\thinspace d. A secondary modulation with much longer period is also suspected in this star. In other two stars the modulation is suspected based on time-dependent analysis of the merged data, in which cyclic phase changes are apparent (amplitude changes are also present, but not as pronounced). Because of a different time-scale of modulation, the origin of the phenomenon might be different in these seven stars with long modulation periods. 

\begin{table*}
\centering
\caption{Cepheids with a likely long-period modulation. Results based on the analysis of merged OGLE-III and OGLE-IV data. The columns contain: star's id, pulsation period, estimate of the modulation period and remarks.}
\label{tab:longp}
\begin{tabular}{lrrl}
id & $\Pf$ (d) & $\Pm$ (d) & remarks \\
\hline
SMC-0423 & 12.41185(3) & 2835(29) & peaks at $\fF\pm\fm$, $2\fF+\fm$ and $\fm$; suspected second modulation with $\Pm>6500$\thinspace d\\
SMC-0208 & 16.68077(4) & $\sim1100$ & complex structures at $k\fF$ and $\fm$; periodic phase changes\\
SMC-0359 & 17.46240(2) & 707(4)    & symmetric side peaks at $\fF$ are mutual 1-yr aliases\\
SMC-1403 & 28.7820(1)  & $>5000$  & unresolved power at $k\fF$, smooth and non-linear phase change\\
\hline
\end{tabular}
\end{table*}

In two stars from the SMC (OGLE-SMC-CEP-0387 and OGLE-SMC-CEP-1065) and in one star from the LMC (OGLE-LMC-CEP-2008) we detect two modulation periods. We also report a few candidates in which secondary modulation is possible (`sm' in the last column of Tabs.~\ref{tab:smc} and \ref{tab:lmc}). OGLE-SMC-CEP-1065 has a very strong modulation with $\Pm=141.4$\thinspace d which we illustrated in Fig.~\ref{fig:lcSMC}. The secondary modulation is much weaker. Two side peaks were detected on positive side of $\fF$ and $2\fF$ with amplitudes below $4$\thinspace mmag. Another interesting case is OGLE-LMC-CEP-2008 in which two modulation periods are in resonance, $P_{\rm m1}/P_{\rm m2}=3\!:\!2$. In the low-frequency range of the power spectrum, we detect peaks at linear combinations of the two modulation frequencies (see note `B' in Tab.~\ref{tab:lmc}). Finally, the very short primary modulation in OGLE-SMC-CEP-0387, may also be interpreted as due to excitation of the radial first overtone. We discuss this interesting case in Sect.~\ref{ssect:rezodm}.

\subsection{Amplitude of the modulation}\label{ssect:amps}
%%%%%%%%%%%%%%%%%%%%%%%%%%%%%%%%%%%%%%%%%%%%%%%%%%%%%%%%%%

Additional peaks in the frequency spectrum, which we relate to modulation, are always in the mmag range. As already noted, in the majority of cases the effect is detected only thanks to the Fourier analysis of the data. A striking feature of nearly all stars (25 out of 29 in the SMC and 18 out of 22 in the LMC), is the presence of the peak at the modulation frequency, $\fm$, which, in many stars, is the highest among the modulation peaks, or is comparable to the side peaks at the pulsation frequency. It is well visible in Tabs.~\ref{tab:smc} and \ref{tab:lmc}, in which the highest among the modulation peaks is marked with bold-face font, and in Figs.~\ref{fig:fspSMC} and \ref{fig:fspLMC} showing the frequency spectra for selected stars. In a few stars, the peak at the harmonic, $2\fm$, is detected as well (see notes on frequency content in the last column of Tabs.~\ref{tab:smc} and \ref{tab:lmc}). The distinct presence of the peak at the modulation frequency means, that the mean brightness of Cepheids is modulated, a fact of extreme importance for the use of these stars as standard candles -- we will discuss this problem in Sect.~\ref{ssect:candles}. 

\begin{figure}
\includegraphics[width=\columnwidth]{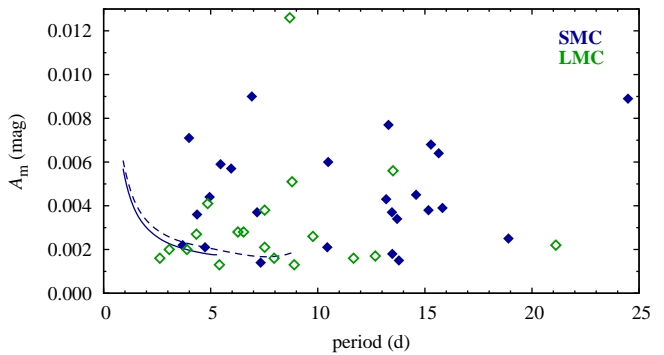}
\caption{Amplitude of the mean brightness modulation, $A_{\rm m}$, versus the pulsation period. Solid and dashed lines are the detection limit curves for the SMC copied from Fig.~\ref{fig:noise} (mean $I$-band brightness was transformed to pulsation period by fitting a linear function to $I$ vs. $\log P$ relation).}
\label{fig:amvsp}
\end{figure}

\begin{figure}
\includegraphics[width=\columnwidth]{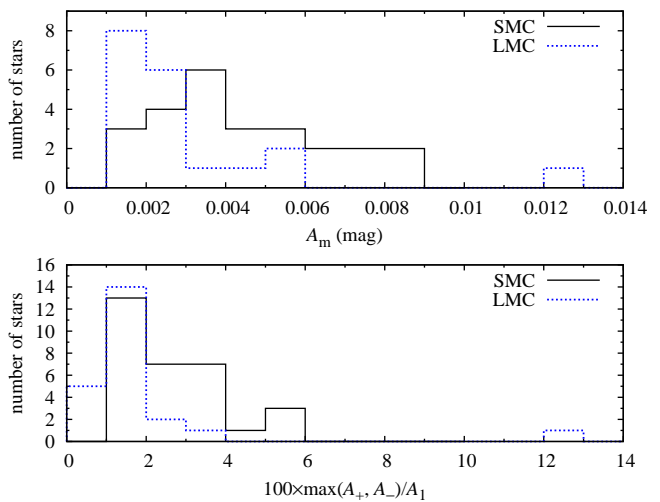}
\caption{Distribution of the amplitudes of mean brightness modulation (top) and of the relative modulation amplitude of the fundamental mode (bottom).}
\label{fig:histo_amps}
\end{figure}

Amplitudes of the modulation peaks are different for stars of the two Magellanic Clouds. The effect is clearly weaker in the LMC, as comparison of Tabs.~\ref{tab:smc} and \ref{tab:lmc} indicates. This is also visible in Fig.~\ref{fig:amvsp} in which we plot the amplitude of the peak detected at $\fm$, $A_{\rm m}$, which corresponds to the amplitude of mean brightness modulation, versus the pulsation period. We observe no obvious dependence on the pulsation period. In Fig.~\ref{fig:histo_amps}, we show the histogram of amplitudes of the mean brightness modulation. With the exception of OGLE-LMC-CEP-1647, the peaks at $\fm$ are always below $0.01$\thinspace mag and are typically lower in the LMC than in the SMC. The histogram of relative modulation amplitude of the radial mode, i.e. of $\max(A_+,A_-)/A_1$, in which $A_-$ and $A_+$ refer to the amplitudes of the side peaks at $\fF-\fm$ and $\fF+\fm$, respectively, is presented in the bottom panel of Fig.~\ref{fig:histo_amps}. The record holder is OGLE-LMC-CEP-3574; its nearly $13$\thinspace per cent relative modulation amplitude results from its small pulsation amplitude (expected for Cepheids close to the 2:1 resonance centre, see Sect.~\ref{ssect:modulatedQM}). In all other cases the relative modulation amplitude is below $6$\thinspace per cent, with systematically lower values in the case of LMC. 

\begin{figure}
\includegraphics[width=\columnwidth]{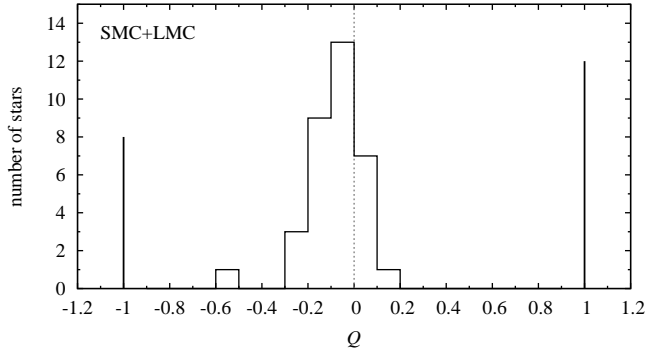}
\caption{Distribution of the asymmetry parameter, $Q$, for all stars form the LMC and SMC. Vertical bars at $|Q|=1$ indicate the number of stars in which only one side peak was detected at radial mode frequency.}
\label{fig:q}
\end{figure}

The other interesting feature of the analysed sample emerges when amplitudes of the side peaks at radial mode frequency are considered. If two side peaks are detected, then their amplitudes are comparable in nearly all the cases. We illustrate it in Fig.~\ref{fig:q}, in which we plot the histogram of the asymmetry parameter defined as: $Q=(A_+-A_-)/(A_++A_-)$ \citep{alcock_rrab}. Since the shape of the distributions turned to be qualitatively the same for LMC and SMC, we only plot the cumulative distribution for all stars of both Clouds. For side peaks of the same amplitude $Q\!=\!0$, and for highly non-symmetric cases $|Q|$ approaches unity. We see a very strong preference towards symmetry, with lower frequency side peak typically slightly higher than the higher frequency side peak. In all cases $|Q|<0.3$, with the exception of OGLE-SMC-CEP-0896, in which $Q\!\approx\!-0.5$ (we note that this star has the shortest modulation period in our sample). On the other hand, in several stars we detect only one side peak, either a lower frequency one, or a higher frequency one. This is indicated with vertical bars at $Q=\pm 1$ in Fig.~\ref{fig:q}. We note that among stars in which only one side peak was detected, strong detections wit $\sn$ well above 5 are not rare. Hence, the scenario in which both side peaks are present in the star, and are of similar amplitude, but one of them remains insignificant due to e.g. random noise fluctuation, is unlikely. Our sample seems to divide into two subsamples: with nearly symmetric side peaks at radial mode frequency and with only one side peak, either on positive side, or on negative side of the radial mode frequency. We will discuss the possible origin of this dichotomy in Sect.~\ref{ssect:modulatedQM}. Here we stress, that the stars of two groups are no different when their other properties are considered, e.g. location in the colour-magnitude and period-luminosity diagrams, light curve shape, or modulation periods. We note however, that stars with only one side peak at the radial mode frequency, prefer longer pulsation periods, at least in the SMC, in which all have $\Pf>10$\thinspace d.

The pattern of the frequency spectrum seems to gradually change as pulsation period increases. In Tabs.~\ref{tab:smc} and \ref{tab:lmc}, the bold face font marks the amplitude of the highest side peak. Systematics is clear in the SMC (Tab.~\ref{tab:smc}). For periods below $\approx 10$\thinspace d, it is the low frequency peak at $\fm$, that is typically the highest. In a few short period stars the peak at $\fF-\fm$ is higher, but in these cases the peak at $\fm$ is of comparable amplitude. As pulsation period increases, the peak at $\fF-\fm$ starts to dominate and finally, the peak at $\fF+\fm$ takes over. As mentioned, there are exceptions, but the overall trend is clear. In the LMC (Tab.~\ref{tab:lmc}) the same trend is observed but is less pronounced. Although in three stars with the shortest pulsation period the peak at $\fF-\fm$ is the highest, the one at $\fm$ is of comparable height. Previously, we have noted that the modulation period also changes with the pulsation period; a slight trend of increasing modulation period with increasing pulsation period was noted (Fig.~\ref{fig:periods}).

Finally, we comment on the lack of modulated stars among the fainter, short-period Cepheids in the SMC, well visible in Figs.~\ref{fig:pl} and \ref{fig:cmd}. Modulated stars were not detected in Cepheids with $\Pf\lesssim 3.6$\thinspace d, or alternatively in Cepheids fainter than $I\approx15.5$\thinspace mag. This is not an observational selection effect, which we illustrate with Fig.~\ref{fig:amvsp}, in which we overplotted the detection limits for the SMC from Fig.~\ref{fig:noise}. In Figs.~\ref{fig:amvsp} (and in Fig.~\ref{fig:noise}) we see that the noise level in the Fourier transform increases as pulsation period (mean brightness) decreases. The increase at $\Pf\approx 3.6$\thinspace d ($I\approx 15.5$\thinspace mag) is not sharp however. The signals of $3-4$\thinspace mmag should be easily detected for pulsation periods $\Pf\gtrsim 2$\thinspace d (up to $I=16.5-17$\thinspace mmag), so in the majority of the SMC Cepheids. We conclude that the lack of fainter, short-period Cepheids in the SMC is not an observation selection effect. The modulation must disappear for periods shorter than $\Pf\approx 3.6$\thinspace d or amplitude of the effect must drop sharply then. We note that in the shortest-period modulated Cepheid in the SMC, the amplitudes of the modulation side peaks are indeed very small ($2$\thinspace mmag), but the number of short-period modulated stars is too low to conclude about sharp weakening of the modulation. The period limit for modulated stars might be connected to the mechanism behind the modulation, as we comment in the Discussion. 

\subsection{The SMC versus the LMC}\label{ssect:witd}
%%%%%%%%%%%%%%%%%%%%%%%%%%%%%%%%%%%%%%%%%%%%%%%%%%%%%%

Properties of the modulation differ in the two Magellanic Clouds. The most striking difference is the strength of the effect, which is significantly lower in the LMC: amplitudes of the modulation side peaks are systematically lower. Also, in no period range the effect is frequent in the LMC, while in the SMC, the incidence rate may be as high as $37$\thinspace per cent for $12<\Pf<16$\thinspace d.

Metallicity seems the most obvious factor that could be responsible for these differences. The two Magellanic Clouds differ in metal content, LMC being more metal rich. Most of the metallicity measurements are for the old population of RR~Lyr stars for which determinations are done with the photometric method \citep[see][]{jk96}. \cite{haschke} estimated the mean metallicity of the Clouds to $-1.50\pm0.24$ and $-1.70\pm0.27$, for the LMC and SMC, respectively (on the metallicity scale of Zinn \& West). It indicates that lower metallicity favours the modulation. Metallicity also affects the position and strength of the 2:1 resonance, which, as discussed below, might be the mechanism responsible for the modulation.

\section{Discussion}
%%%%%%%%%%%%%%%%%%%%

\subsection{Is it the genuine modulation?}\label{ssect:modulatedQM}
%%%%%%%%%%%%%%%%%%%%%%%%%%%%%%%%%%%%%%%%%%%%%%%%%%%%%%%%%%%%%%%%%%%

In the frequency spectrum, the modulation of pulsation amplitude and/or phase manifests as equally spaced multiplet structures centred at $k\fF$. In addition, a signal at the modulation frequency, $\fm$, might be detected, which corresponds to modulation of the mean brightness. This is the case for RR~Lyr stars in which the modulation of pulsation amplitude and/or phase is common and known as the Blazhko effect \citep[see e.g.][]{geza,s16bl}. However, the signal at $\fm$ is rarely detected in RR~Lyr stars, mostly in the top-quality photometry, and is of amplitude much lower than amplitude of the side peaks at $k\fF$ \citep[see e.g.][]{benko14}. In the frequency spectra of the discussed Cepheids, except the side peaks at $k\fF$, strong peak at the modulation frequency, $\fm$, is detected in the majority of cases. In stars with pulsation periods below $\approx\!10$\thinspace d, this peak is typically the highest among the modulation peaks. 

\begin{figure}
\includegraphics[width=\columnwidth]{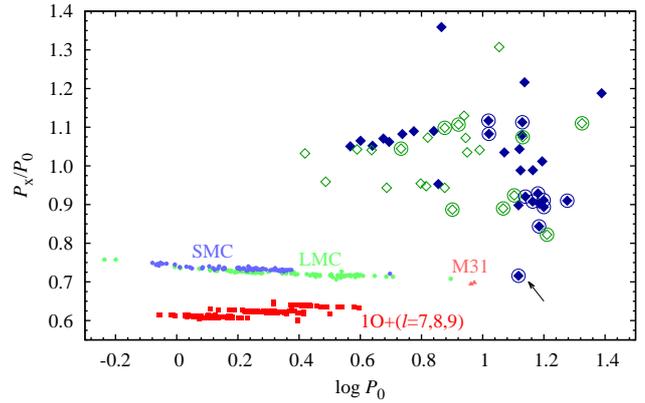}
\caption{Petersen diagram for the discussed Cepheids, constructed assuming that the higher side peak at radial mode frequency corresponds to non-radial mode with period $P_{\rm x}$. Stars in which only one side peak was detected at radial mode frequency are encircled. For comparison, double-mode fundamental and first overtone Cepheids from the OGLE collection \citep{o4_mc_cep} are plotted with small dots. Two long-period, double-mode Cepheids from M31 \citep{poleskiM31} are plotted with small triangles. Double-mode first overtone and non-radial, $\ell=7,8$ and $9$ Cepheids \citep{ss16} are plotted with small squares.}
\label{fig:pet}
\end{figure}

In principle, one of the peaks, which are now collectively interpreted as due to modulation, may correspond to non-radial pulsation mode. Other peaks are linear frequency combinations then (expected for the double-mode pulsator), as illustrated in Tab.~\ref{tab:scenarios}. In Fig.~\ref{fig:pet}, we plot the Petersen diagram for the scenario in which one of the side peaks at radial mode frequency (the one of higher amplitude) corresponds to additional pulsation mode. On vertical axis we plot the $P_{\rm x}/\Pf$ ratio, so for side peaks of lower frequency the ratio is larger than $1$. For comparison, a tight sequence of double-mode Cepheids pulsating simultaneously in the fundamental and first overtone modes from the OGLE collection is plotted ($P_1/P_0$ vs. $\log P_0$). Also, double-mode Cepheids that pulsate simultaneously in the radial first overtone and in acoustic, non-radial modes of $\ell=7,8$ and $9$, are plotted with small red squares \citep[$P_{\rm x}/P_1$ vs. $\log P_1$;][]{ss16,wd16}. These stars form three well defined and narrow sequences in the diagram. To the contrary, for the discussed Cepheids we observe no groupings or tight progressions that could correspond to excitation of specific non-radial mode(s). This is true both for the stars in which two side peaks are detected at the radial mode frequency, and for the stars in which only one side peak was detected. The latter stars are encircled in Fig.~\ref{fig:pet}.

Theory of stellar pulsation also indicates that interpretation in terms of non-radial modes is unlikely. For the typically highest peaks at $\fm$, or at $\fF-\fm$, the only possibility are non-radial modes of gravity or of mixed character (as these frequencies are lower than the frequency of the radial fundamental mode, $\fF$). Excitation of gravity modes is not expected in these giant stars, see \cite{wd77}. The stability of non-radial acoustic modes in RR~Lyr-type models was studied by \cite{vdk98}, but their conclusions are relevant also for other giants in the classical instability strip. They found that a large number of unstable, low-degree, $\ell=1-2$ modes, have frequencies close to the frequencies of the radial modes. Thus, for the peak at $\fF+\fm$, non-radial acoustic pulsation is also, in principle, possible. This peak is rarely the highest one, however (see Tabs.~\ref{tab:smc} and \ref{tab:lmc}). Also, theoretically predicted growth-rates of these non-radial modes are significantly smaller than growth rates of the radial modes. Their spectrum is dense and there is no obvious mode selection mechanism. Hence, we rule out the possibility that detected peaks are due to non-radial pulsation. 

Additional argument in favour of modulation is the detection of equally-spaced multiplets (or their components) in some Cepheids, although these cases are rare. These are OGLE-SMC-CEP-3490, OGLE-SMC-CEP-2574, OGLE-LMC-CEP-1083 and OGLE-LMC-CEP-0139 -- see the last columns of Tabs.~\ref{tab:smc} and \ref{tab:lmc}.

Periodic modulation may be caused by the presence of a spot and stellar rotation. We note however, that for the majority of stars, the modulation peaks are coherent and their frequencies and amplitudes remain more or less constant over the OGLE-III and OGLE-IV observations -- for more than 15 years. Hence, transient phenomena, like spots, do not seem a good explanation. 

Genuine modulation seems the most realistic explanation, although its properties are clearly different from the Blazhko modulation detected in RR~Lyr stars. Based on the properties of the frequency spectra, Cepheids seem to divide into two groups -- those with symmetric side peaks centred at radial mode frequency (and its harmonics), and those with one of the side peaks missing. In both groups, the peak at modulation frequency is commonly present. For the former group (symmetric side peaks), we may conclude that amplitude modulation is dominant \citep[see][]{benko11}. For the latter group (one of the side peaks not detected), modulation scenario implies that both side peaks are present, but are extremely non-symmetric (and hence, one of them is not detected). Physically, such pattern is produced when combined amplitude and phase modulation of comparable strength take place. In addition, a close to $\pm\uppi/2$ phase lag between the amplitude and phase modulations is necessary \citep{benko11}. As noted in Sect.~\ref{ssect:amps}, Cepheids with only one side peak detected, tend to have longer pulsation periods.

The properties of modulation are a consequence of the mechanism at work, which is unknown. We speculate however, that modulation may be caused by the 2:1 resonance between the fundamental mode and the second overtone, $2\fF\!=\!\nu_{2}$, the resonance that shapes the Cepheid bump progression \citep[e.g.][]{ss76,bk86,bmk90}. Arguments are the following: (i) The majority of modulated stars are bump Cepheids, with clearly marked bump feature. Only for the shortest pulsation periods, the bump is not present, but it is clear that the 2:1 resonance shapes the Cepheid light curves in period range much wider than determined solely from the occurrence of the bump -- see the Fourier parameter plots in Fig.~\ref{fig:fp}; (ii) Modulation properties of the Cepheids, i.e. modulation period and amplitudes of the side peaks, change with the increasing pulsation period, as discussed in the previous Section. The light curve shape of the bump Cepheids also evolves with the pulsation period. In fact, it is the distance from the resonance centre, which we may define as $\Delta\!=\!P_2/P_0\!-\!0.5$, that determines the light curve shape \citep[e.g.][]{bmk90}. It might also be the parameter that affects the modulation properties of Cepheids. In particular, the effect might be limited to stars sufficiently close to the resonance centre, which would explain the lack of modulation among short-period SMC Cepheids. Although it might seem that modulated star with extreme period, e.g. with period above $20$\thinspace d, is far from the resonance centre, the value of $\Delta$ might be small also in this case; it depends on physical parameters of the star, see fig.~3 in \cite{bmk90}. Theoretical modelling indicates that the Fourier phase, $\varphi_{21}$, is tightly related to $\Delta$ -- a fact that might facilitate further comparison with the theory.

So far, modulation was not reproduced in non-linear models of fundamental mode Cepheids. A wide survey of convective models is needed however, as most of the modelling published so far was done with purely radiative codes. Interestingly,  modulation was found in the convective models of type~II Cepheids, in \cite{sm12} and \cite{s16}. In these models, the mean stellar parameters are modulated, just as we observe in the majority of classical Cepheids discussed here -- see the frequency spectra in fig.~2 in \cite{sm12} and in fig.~10 in \cite{s16}, all with prominent peak at the modulation frequency.

While the search for modulation with non-linear calculations is time-consuming and there is no guarantee of succeeding (the codes that can be used in such massive computations are 1D with simplified treatment of convection), the other theoretical tool is available. With the help of amplitude equation formalism \citep[e.g.][]{bk86}, it can be checked whether the proposed mechanism is feasible. The results of \cite{pam86}, who studied the resonances as a possible explanation of the Blazhko effect in RR~Lyr stars, are promising in this context. Based on the analysis of amplitude equations, they pointed that the 2:1 resonance between the fundamental mode and a higher-order radial overtone may, in certain conditions, cause the periodic modulation of pulsation. We note that for RR~Lyr stars, this mechanism cannot work, as the proposed resonance (with the radial third overtone) was pushed far beyond the instability strip after the opacity tables were revised. Detailed analysis of the amplitude equations is beyond the scope of the present paper; dedicated study is planned.

\begin{table}
\centering
\caption{Interpretation of the frequency spectrum peaks in different scenarios: modulation with frequency $\fm$, and beating with non-radial mode of frequency $\nu_{\rm x}$.}
\label{tab:scenarios}
\begin{tabular}{rrrr}
scenario                 & \multicolumn{3}{c}{peak interpretation}\\
\hline
periodic modulation      &   $\fm$   & $\fF-\fm$  & $\fF+\fm$  \\
beating 1 (gravity mode) &   $\fx$   & $\fF-\fx$  & $\fF+\fx$  \\
beating 2 (gravity mode) & $\fF-\fx$ & $\fx$      & $2\fF-\fx$ \\
beating 3 (acoustic mode)& $\fx-\fF$ & $2\fF-\fx$ & $\fx$      \\
\hline
\end{tabular}
\end{table}

\subsection{OGLE-SMC-CEP-0387: Extreme resonant double-mode pulsator?}\label{ssect:rezodm}
%%%%%%%%%%%%%%%%%%%%%%%%%%%%%%%%%%%%%%%%%%%%%%%%%%%%%%%%%%%%%%%%%%%%%%%%%%%%%

OGLE-SMC-CEP-0387 deserves special attention. This star is listed in Tab.~\ref{tab:smc} with two modulation periods. If the highest side peak at radial mode frequency (at $\fF+\fm$), associated with the primary modulation, is interpreted as due to additional mode, then, the corresponding point in the Petersen diagram is located at the long-period extension of the F+1O double-mode Cepheid sequence (it is marked with arrow in Fig.~\ref{fig:pet}). Although its fundamental mode period is long ($\Pf\approx13.1$\thinspace d) and the period ratio, $0.716$, is a bit too high to match the extension of the F+1O double-mode SMC sequence exactly, the scenario in which the peak now associated with modulation is in fact radial first overtone cannot be rejected. We note that \cite{poleskiM31} discovered two double-periodic, long-period Cepheids in M31 and interpreted them as double-mode F+1O Cepheids. These stars are marked in Fig.~\ref{fig:pet} with small triangles. In fact, long-period double-mode pulsation caused by the 2:1 resonance between the fundamental mode and the second overtone, that we have just discussed, was predicted by \cite{dk84}, based on analysis of the amplitude equations. Such double-mode pulsation is expected close to the resonance centre, at fundamental mode periods not far from $10$\thinspace d. Resonant double-mode pulsation was reproduced in non-linear pulsation models, very close to the resonance centre -- see \cite{eas} and \cite{buchler09}. Whether such form of pulsation can extend to periods as high as $13$\thinspace d needs to be verified with non-linear calculations. The possibility is appealing, but the genuine modulation is likely as well.

\subsection{Standard candles with a flaw?}\label{ssect:candles}
%%%%%%%%%%%%%%%%%%%%%%%%%%%%%%%%%%%%%%%%%%%%%%%%%%%%%%%%%%%%%%%

Strong peak at $\fm$ indicates, that mean brightness of the Cepheid is modulated. Consequently, the apparent brightness changes over time and so is the distance modulus calculated as the difference between the measured, apparent brightness and absolute brightness computed through the $P-L$ relation. For studies using large sample of Cepheids with numerous observations gathered over years, e.g. studies of the structure of the Magellanic Clouds using the data gathered by photometric sky surveys, like OGLE, the effect can be safely ignored. Reasons are the following: (i) Amplitude of the mean brightness modulation is small, typically well below $0.01$\thinspace mag (Fig.~\ref{fig:histo_amps}, top panel; we stress however, that through this paper we use Fourier amplitudes, so the peak-to-peak variation is roughly two times larger); (ii) The effect averages out if the light curve was gathered over longer time base; (iii) Except for specific fundamental mode period ranges, the effect is very rare (Fig.~\ref{fig:histo_ic}). 

For the same reasons, the modulation has negligible effect on the use of Cepheids in constructing the cosmological distance ladder. We note that modulation might be weaker in more metal rich systems, as comparison of the LMC and SMC suggests (Sect.~\ref{ssect:witd}). Most of the galaxies used in the calibration of Cepheid distance scale are young, metal rich spirals \citep[for a review see][]{fm10}. Direct detection of modulation in typical observations of Cepheids in distant galaxies is unlikely, as the number of gathered epochs is usually small and the photometric accuracy is low. Modulation, if indeed present in the Cepheid, will contribute to the minor increase of the light curve scatter. If only a single pulsation cycle is covered, the mean apparent brightness might be shifted, but by no more than a few mmag. The common approach used in calibration of any standard candles, i.e. observation of sufficiently abundant sample in order to reduce the associated statistical uncertainties, should mitigate the effect of possible modulation in Cepheids, which, in any case, is very small.

\section{Summary and conclusions}
%%%%%%%%%%%%%%%%%%%%%%%%%%%%%%%%%

We have discovered low amplitude, periodic modulation of pulsation in 29 fundamental mode Cepheids in the SMC and in 22 fundamental mode Cepheids in the LMC. The relative modulation amplitude is low. In all but one case, it is below $6$\thinspace per cent. In nearly all stars, modulation of the mean brightness is detected, which manifests through the presence of prominent peak at the modulation frequency in the frequency spectra of the stars. Amplitude of this modulation is also low. In all but one star, it is below $0.01$\thinspace mag. In the record holder, OGLE-LMC-CEP-1647, amplitude of this peak is $0.013$\thinspace mag. The effect is systematically weaker in the LMC, pointing that it may be metallicity dependent. Modulation periods vary from 30 to more than 1000 d, but in the majority of cases the typical modulation timescale is ten times longer than the pulsation period (between $\sim\!70$ and $\sim\!300$\thinspace d).

The effect of modulation on the distance determination, using the Cepheids' period-luminosity relation may be safely ignored. Modulation is rare, of low amplitude and averages out in typical observations gathered over long time base.

Although the overall incidence rate is very low, of order of $1$\thinspace per cent in both Magellanic Clouds, the effect may be more common in specific ranges of the pulsation period. In particular, in the SMC, the incidence rate is $37$\thinspace per cent for Cepheids with pulsation periods between $12$ and $16$\thinspace d. For the LMC, the largest incidence rate is of order of $5$\thinspace per cent for fundamental mode periods between $8$ and $14$\thinspace d. Modulation was not detected in any star with pulsation period below $2.6$\thinspace d. Although in the LMC such short period fundamental mode Cepheids are rather rare ($\approx\!10$\thinspace per cent of the OGLE sample), in the SMC they constitute the majority of the OGLE sample ($\approx\!65$\thinspace per cent). 

Modulation in fundamental mode Cepheids may be a more common phenomenon, but because of its low amplitude, its detection in the ground-based photometry is challenging. In several stars we have detected a single peak close to the radial mode frequency or at the low frequency range. Such single periodicity may be due to contamination, but may also indicate a possible modulation. We note that periodic modulation was detected in V1154~Cyg, fundamental mode Cepheid ($\Pf\approx4.925$\thinspace d) observed with {\it Kepler}/{\it K2} \citep{kanev,derekas17}. Modulation period is $\Pm\approx 159$\thinspace d, but amplitudes of the modulation side peaks are nearly an order of magnitude lower than reported here; the highest side peak, at $\fF-\fm$, has an amplitude of $0.55$\thinspace mmag. Just as for the Cepheids in the Magellanic Clouds, the peak at modulation frequency is relatively high, $0.37$\thinspace mmag.

Observations with the photometric space telescopes will be crucial to establish the incidence rate of the light curve modulation in classical Cepheids. The brightest Cepheids in the sky will be observed with {\it TESS} [\cite{TESS}; for prospects of Cepheid's observations see \cite{molnarTESS}], but for the majority of them, observations will be limited to short, 27d-long observing runs -- too short to resolve the modulation. Situation will be much better for stars located within and close to the Continuous Viewing Zone, so also for the LMC Cepheids. These are however faint and typically located in the crowded fields -- only for the brightest Cepheids we can expect photometry allowing to detect the mmag signals.

The origin of modulation remains unknown, however, we speculate that it may be a resonant effect, caused by the 2:1 resonance between the fundamental mode and the second overtone. This resonance shapes the Cepheid bump progression. In fact, majority of the modulated stars are bump Cepheids, with prominent bump feature in their light curves. In the SMC, the effect is frequent close to the resonance centre. Properties of the modulation vary as pulsation period increases, just as light curve shape evolves due to the resonance. A thorough study of the relevant amplitude equations and of non-linear pulsation models is planned to verify the resonance hypothesis.

Quasi-periodic modulation of pulsation is common in RR~Lyr stars, in which it is known as the Blazhko effect. In classical Cepheids, the modulation was also reported, but so far in a rather specific double-overtone stars, in a few first overtone pulsators and in the oddball, second overtone star, V473~Lyr \citep[see e.g.,][respectively]{mk09,o4_mc_cep_pec,ms14}. Here, we have reported the discovery of periodic modulation in the sizeable sample of long-period fundamental mode stars, commonly used as distance indicators. Although the modulation has no practical effect on the use of Cepheids as standard candles, it demonstrates that our understanding of these important classical variables is far from being complete.

\section*{Acknowledgements}
This research is supported by the Polish National Science Centre, grant agreement DEC-2015/17/B/ST9/03421. I am grateful to Pawel Moskalik for thorough reading of the manuscript and important comments.

%%%%%%%%%%%%%%%%%%%%%%%%%%%%%%%%%%%%%%%%%%%%%%%%%%



%\bibliographystyle{mnras}


%%%%%%%%%%%%%%%%%%%%%%%%%%%%%%%%%%%%%%%%%%%%%%%%%%

%%%%%%%%%%%%%%%%% APPENDICES %%%%%%%%%%%%%%%%%%%%%

%\appendix

%\section{Additional figures, not intended for publication, including on-line}

%If you want to present additional material which would interrupt the flow of the main paper, it can be placed in an Appendix which appears after the list of references.

%%%%%%%%%%%%%%%%%%%%%%%%%%%%%%%%%%%%%%%%%%%%%%%%%%

% Don't change these lines
\bsp	% typesetting comment
\label{lastpage}
\end{document}